\documentclass[12pt]{article}

\usepackage{natbib}

\usepackage[dvips]{epsfig}

\begin{document}

\title{Compression ratios based on the Universal Similarity Metric  still yield protein distances far from CATH distances}

\author{ F. Rossell\'o  \hspace*{1cm} J. Rocha \hspace*{1cm} J. Segura  \hspace*{1cm} \\[1ex]
{\small Dept.\ of Mathematics and Computer Science,}\\
{\small University of the Balearic Islands,}\\
{\small 07122 Palma de Mallorca (Spain)}\\
{\small \emph{E-mail:}
\{\texttt{jairo,cesc.rossello}\}\texttt{@uib.es}
}
}

\maketitle

\begin{abstract}
Kolmogorov complexity has inspired several alignment-free distance
measures, based on the comparison of lengths of compressions, which
have been applied successfully in many areas.  One of these measures,
the so-called Universal Similarity Metric (USM), has been used by
Krasnogor and Pelta to compare simple protein contact maps, showing
that it yielded good clustering on four small datasets.  We report an
extensive test of this metric using a much larger and representative
protein dataset: the domain dataset used by Sierk and Pearson to
evaluate seven protein structure comparison methods and two protein
sequence comparison methods.  One result is that Krasnogor-Pelta
method has less domain discriminant power than any one of the methods
considered by Sierk and Pearson when using these simple contact maps.
In another test, we found that the USM based distance has low
agreement with the CATH tree structure for the same benchmark of Sierk
and Pearson.  In any case, its agreement is lower than the one of a
standard sequential alignment method, SSEARCH. Finally, we manually
found lots of small subsets of the database that are better clustered
using SSEARCH than USM, to confirm that Krasnogor-Pelta's conclusions
were based on datasets that were too small.
\end{abstract}

\section{Introduction}

\citet{Pelta1}  represented a protein structure by means of
the list of contacts (position pairs) and used a compress algorithm
to approximate the Kolmogorov complexity on which the definition
of the Universal Similarity Metric is based.  To test the metric they applied it independently to
four protein datasets comprising 122 proteins and clustered
each set using an off-the-shelf hierarchical clustering software
applied to the resulting inter-distance matrices.  In each case they
obtained an almost correct classification of the proteins.
They inferred that the Universal Similarity Metric captures protein
structure similarity.

Since each aminoacid  has its own sequential position in a protein, and the 
contact pairs are  pairs of these positions, we were surprised that the method
could find similarities between  two different proteins each with its own numbering.
But the authors conclude  the following in the discussion section of the above mentioned paper:
``This new metric can be used as a robust measure of similarity of domains where
either, there is no enough modeling information, or there is no consensus
on which aspects are to be modeled. 
We gave mathematical and experimental evidence that USM can be used to successfully
assess protein structure similarity. The USM seems to be capable of 
capturing protein similarities  that encompasses  a variety of other,
more heuristic, criteria on a fully automated way. It seems that the
Universal Similarity Metric is so robust that even with a rough guess
of parameters it is still possible to deliver good results. One disadvantage 
of using USM on its own is that, although it can differentiate between protein families and sub-families and measure similarity based on a
rigorous mathematical definition, it does not give indications of where
these (di)similarities come from.'' Given these strong conclusions
on the robustness of their method to differentiate between protein sub-families, first, we downloaded their algorithms and data-sets and 
confirmed exactly their results using this simple protein format. Then,
we set off for verifying the results on a bigger set because if they
are satisfactory we could abandon our search for sophisticated comparison
methods.

We present in this paper a bigger test on almost 2800 protein domains
already used to assess the performance of the most important
algorithms used nowadays for protein comparison, namely the 
benchmark prepared by 
\citet{Sierk}.  
When we apply the
same Sierk-Pearson test, a though one, that was applied to the other comparison algorithms,
 the method revels to be
useless for helping to decide whether two proteins are similar or not;
 the result is that the Universal Similarity
Metric using this contact map format and compression approximation performs quite worse than any method considered by
Sierk and Pearson, even the ones that only use as input the protein sequences. Therefore, the method as it stands  does not encompasses other criteria for comparison as the authors claimed. 

 Therefore, a global analysis of the method is needed to find
out under which conditions it is reliable,  if any. One way to undertake this
analysis is to compare the matrix generated by USM for the domains in the database used above with respect to
the metrics associated to the CATH tree that represents the family relationships
of the domains. We found the distance on the tree that best approximates the
USM matrix.  Then, we calculate the correlation between the USM matrix and
the tree distance. A low correlation means that the USM matrix does not
represent correctly the tree structure. As we explain in the Results Section
below, we found higher correlations for the distance generated by SSEARCH 
\citep{SSEARCH} than
to for USM. In short, we have a global evidence that clusterings produced
by USM are farther from the CATH tree than the ones produced by a sequential
method that does not uses 3D information.

We also confirm manually this result.
We found several (more than ten) easily classifiable datasets contained in the benchmark set where the
proposed method does not classify correctly the domains. Since the
Krasnagor-Pelta method only has three small real datasets as evidence to argue that
it  can be used to successfully assess protein structural similarity but 
we can find several easy subsets in the  benchmark  where  it performs poorly, 
the robustness of the method is open to doubt. We have found subsets
of proteins in very different families that the method cannot classify 
correctly but SSEARCH can, even thought this one uses 
only sequence information. 

All data and programs used for the preparation of this paper are
available upon request.  See {\tt dmi.uib.es/people/jairo/USM/} for
cluster examples.

In the rest of the introduction, we review the use and notation of the
Kolmogorov complexity on pattern comparison.

As an alternative to general sequence and structure comparison methods based
on alignments, several metrics have been proposed based on Kolmogorov
complexity \citep{BCL02,BGLVZ98,KLR04,Lietal01,Lietal03,VDR03}.
Roughly speaking, the \emph{conditional Kolmogorov complexity}
$K(x|y)$ of two sequences $x,y$ is the length of the shortest binary
program $P$ that computes $x$ with input $y$ \citep{kol65}.  Thus,
$K(x|y)$ represents the minimal amount of information required to
generate $x$ by any effective computation when $y$ is furnished as an
input to the computation.  The \emph{Kolmogorov complexity} $K(x)$ of
a sequence $x$ is defined as $K(x|\lambda)$, where $\lambda$ stands
for the empty sequence.  Given a string $x$, let $x^{*}$ denote the
shortest binary program that produces $x$ on an empty input; if there
are more than one shortest program, we take as $x^{*}$ the first in
alphabetic order.  The \emph{Kolmogorov complexity} $K(x,y)$ of a pair
of objects $x,y$ is the length of the shortest binary program that
produces $x$ and $y$ and a way to tell them apart.  More formal
definitions of all these concepts and their main properties can be
found in the textbook by \citet{livit97}.

The most outstanding metric based on Kolmogorov complexity is the
\emph{Universal Similarity Metric} proposed by 
\citet{Lietal03},
\begin{equation}
d_{u}(x,y)=\frac{\max\{K(x|y^{*}),K(y|x^{*})\}}{\max\{K(x),K(y)\}}.
\label{du}
\end{equation}
These authors proved that this metric (actually, it only satisfies the
axioms of metrics up to a certain additive precision) refines any
other computable similarity metric, like for instance effective
versions of Hamming distance, Euclidean distance, edit distances or
alignment distances \citep[Thm.~VI.2]{Lietal03}.  This Universal
Similarity Metric has been used successfully for instance to compute
phylogenetic trees based on whole mitochondrial genomes
\citep{Lietal03,cilvit05}, cluster SARS virus \citep{cilvit05}, 
classify languages \citep{Lietal03},
musical pieces \citep{cilvitwolf04,cilvit05,LiSleep04}, and images
\citep{svnord04}, detect plagiarism is student assignments
\citep{CFLMS04}, and cluster Russian literature \citep{cilvit05}.
But it has failed to compare TOPS diagrams \citep{TOPSRV}.

Actually, the Universal Similarity Metric was not used in these
applications as it stands, but approximations of it.  The reason is
that Kolmogorov complexities are non-computable in the Turing sense,
and therefore they must be heuristically approximated in practice.
Since $K(x)$ is intuitively the minimal amount of information required
to generate $x$, i.e., the shortest length of a compressed binary
version of $x$, Kolmogorov complexities are approximated by means of
lengths of compressions, and then the formula (\ref{du}) given above
is simplified using suitable properties of Kolmogorov complexity, so
that it no longer involves conditional Kolmogorov complexities.

In this way, and once fixed a compression algorithm, the Kolmogorov
complexity $K(x)$ of an object $x$ is replaced by the length $C(x)$ of
the compression of it using this algorithm.  Furthermore, since
according to \citet{Lietal03}
$$
K(x,y)=K(xy)\mbox{ up to additive logarithmic precision,}
$$
$K(x,y)$ can be replaced by the length $C(xy)$ of a compression of the
concatenation of $x$ and $y$.  Finally, and since
$$
K(x,y)=K(x)+K(y|x^{*})=K(y)+K(x|y^{*})
$$
up to constant additive precision \citep{livit97}, the conditional
complexity $K(x|y^{*})$ can be approximated by $C(xy)-C(y)$, and
$K(y|x^{*})$ can be approximated by $C(xy)-C(x)$ or  by
$C(yx)-C(x)$.

This lead \citet{Lietal03}  to approximate the Universal Similarity
Metric by the \emph{Normalized Information
Distance}
$$
NCD(x,y)=\frac{C(xy)-\min\{C(x),C(y)\}}{\max\{C(x),C(y)\}};
$$
this distance has been thoroughly studied by 
\citet{cilvit05}. A methodological study of its application to
protein sequence classification has been published recently by 
\citet{Kocsoretal05}; they show that a compression based distance combined with
 a BLAST score has a performance  even slightly better than that of the Smith-Waterman algorithm (SSEARCH). 

\citet{Pelta1,Pelta2} have used
a slightly more general approximation of the Universal Similarity Metric
to compare protein structures. The formula they use (and we use in this paper) is
$$
USM(x,y)= \frac{\max \{C(x y) - C(x), C(y x) - C(y)\}} {\max
\{C(x),C(y)\}}.
$$

\section{Methods}

A subset of CATH 2.3 database (which stands for Class, Architecture,
Topology, and Hierarchy \citep{Oregon}) was selected by 
\citet{Sierk} to obtain a non-redundant sample of the entire
database.  A 2771 subset of CATH domains and 86 prototype
domains\footnote{Available at the FASTA repository 
{ftp.virginia.edu:/pub/fasta/prot\_sci\_04}} were selected to test the
following algorithms: Dali, Structal, CE, VAST, Matras, SGM, PRIDE,
SSEARCH and PS-Blast.  This domain set was carefully screened in order
to be considered as a valid benchmark for testing protein comparison
algorithms.  Of the 2771 domains, 1120 belong to the 86 families
of the 86 prototypes.  Therefore, when comparing each of the 2771
domains against each of the 86 prototypes, there is a maximum of 1120
correct hits.  The prototypes are part of these 1120 domains.

For our experiments, the file representation is the one used by Krasnogor
and Pelta: a protein is represented by a list of adjacent aminoacids,
where two aminoacids are adjacent if the distance of the corresponding
$C_\alpha$ atoms are below 6.5 \AA, a threshold also used by them (we also tested
other thresholds as we explain below).
Each line in the file contains two numbers with the first one smaller
than the second one.  The file has a header that consists of two lines: in the
first one, the
number of aminoacids is followed by the comment ``\# Number of Residues''; in the second one, the number of contacts is followed by the comment 
``\# Number of Contacts at 6.5 Angstroms ''. Therefore, two concatenated files could be trivially separated from each other.  The Unix {\tt compress} program was
used.  Other representations of protein structures (binary adjacency
matrix with and without end of line characters) and other compression
programs were tried but are not reported because the results obtained
with them were similar or even slightly worse.

We wanted to carry out two tests: 
The first one is to test its selectivity on the whole database.
The second one is to find out why much agreement there is between the USM based distance among the domains and the tree that CATH associates to them.

With respect to the first test, the best algorithms to  decide whether two proteins are similar have been tested with the 2771 domain
database. Therefore, we  tested if the USM based method 
is useful on deciding this demanding task and compare the results with the
other methods.
Krasnogor-Pelta's {USM} approximation
of the Universal Similarity Metric was applied carrying out a
pairwise comparison of the 86 prototypes versus the 2771 domains.  The
values returned were examined from the best score to the worst.  For
each pair considered in this order whose domains belong to the same
family, a \emph{coverage} value was increased.  Otherwise, an
\emph{error} value was increased.  This is the method used by Sierk
and Pearson to evaluate the other algorithms with the same database.
A perfect classifier would arrive to 100\% coverage before the first
error arrives.

To assess the sensitivity and selectivity of the Universal Similarity
Metric, we plot Errors per Query versus Coverage curves for both
approximations.  These curves show how much coverage is obtained at a
given error level, i.e., the number of true positives detected at a
given number of false positive detected.

With respect to the second test, we aim to compare the hierarchical clusterings
produced by the USM based distance and the fix clustering defined by 
CATH for the same domains. To do so, we find the distance on the tree that
best approximates the USM distance, and calculate the linear correlation
between the two distances. For each tree edge, a weight variable is introduced.
Let $D$ be the domain set and ${\bf x}$ be the vector of edge  weight variables.
We minimize the squares sum of the distances in the tree and in the known
USM matrix:
\[  min \sum_{i,j\in D} (C({\bf x},i,j) - d_{i,j})^2,\]
where  $d_{i,j}$ is the constant USM based distance between domains $i$ and $j$
 and  $C({\bf x},i,j)$ is the sum of the variables associated to the known
edge path between $i$ and $j$ in the CATH tree. Under the positivity 
constrains for the variables, the fit is resolved using standard
 methods for quadratic programming. Finally, the tree distance $t_{i,j}$ 
between the domains $i$ and $j$ is
calculated using the edge weights that produce the best fit. The cophenetic
correlation between $( d_{i,j})$ and $(t_{i,j} )$ is calculated,
and compared to the correlation of the distance generated by SSEARCH.

\section{Results}

\subsection{Selectivity test}

According to  \citet{Pelta1}, their method
sometimes clusters  well proteins that belong to a small number of  groups. 
However, the most important question that needs
to be answered when a new 3D structure of a protein is discovered is 
whether it is similar to one or more better known proteins. The problem of 
clustering proteins is not a crucial one, since lots of structures are not
discovered at once, and clusters are already defined in public databases
using lots of information sources.
\begin{figure}
\begin{center}
\epsfig{file=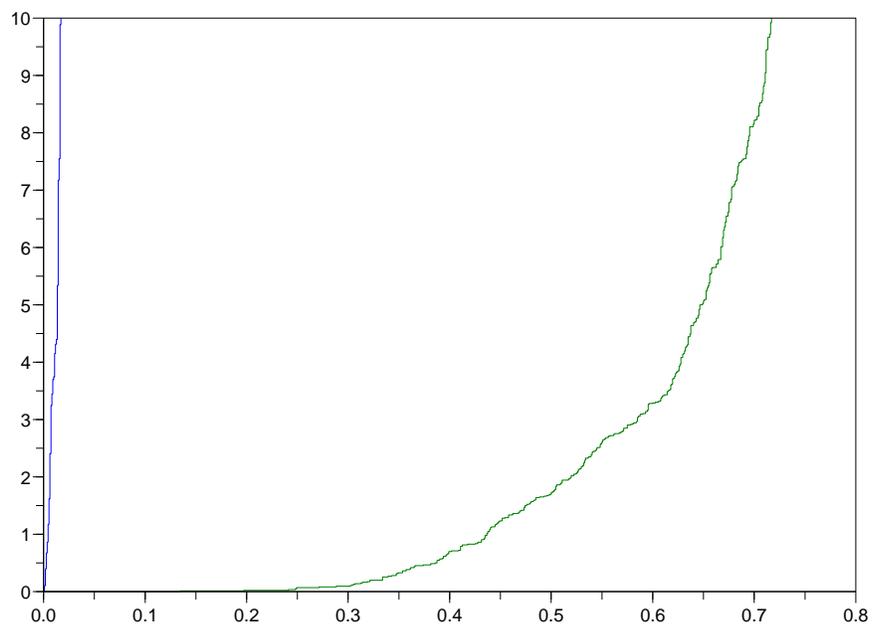, width=\columnwidth}
\end{center}
   \caption{Error-Coverage curves. Left, USM method, right, Dali.}
  \label{ECcurves}
\end{figure}
\begin{table}
\begin{center}
\begin{tabular}{|c|c|c|c|}\hline
{Errors} & USM & {Dali} & {Worst Method} \\
\hline
0.1    &  0.09   & 19 &  10 \\ \hline
1     &  0.4     & 43   &  17 \\ \hline
10    &  1.7   & 71   &  28 \\ \hline
\end{tabular}
\end{center}
\label{errorvscoverage}
\caption{Coverage performace for a given error level for different
protein distances.}
\end{table}

The Error-Coverage curve generated with the Universal Similarity
Metric approximations is quite worse than those corresponding to the
methods considered by Sierk and Pearson, as seen in  Figure \ref{ECcurves}
with respect to Dali.
Table 1 summarizes the results.

The \emph{Errors} column shows the number of errors per prototype
(i.e., 10 corresponds to 860 errors) and all other values are coverage
percentages with respect to 1120 (i.e, 1\% means 11 hits).  The
column {USM}  refers to the results corresponding to the 
approximation explained in the introduction, the Dali column
displays the results corresponding to the structural alignment program
Dali \citep{HS96} and the Worst Method gives the worst result
obtained by any method considered by Sierk and Pearson for the
corresponding error rate.

So, for instance, at 0.1 errors per query on average, i.e., when 8
errors have been found, USM has only correctly covered 0.09\% of the
domains, i.e., 1 domain, while the Dali program at the same error
rate reaches 19\% of correctly covered domains (209 domains) and the
worst method at this error rate (VAST) reaches a coverage of 10\% (112
domains).  At a rate of 1.0 errors per query (86 errors), USM covers
correctly 0.4\% while Dali covers 43\% and the worst method in this case
(SGM) covers 17\%.  At a rate of 10.0 errors per query (860 errors),
the evaluated method covers only 1.7\% while Dali covers 71\% and the
worst method (SGM) covers 28\%.  Notice that his 1.7\% means 19 hits,
i.e., after 860 false positives, USM has not even detected the
identity of all 86 prototypes with themselves.

Other tests, where the contact map is calculated when the threshold
ranges from 
5 to 8 \AA \ gave  the same negative results.

\subsection{Cophenetic correlation}

When comparing a  distance to its closest one on the CATH tree, 
the correlation coefficient for USM is 0.60 and for SSEARCH is 0.87.
This fact  confirms that USM yields 
weaker clusterings with respect to CATH than a method that does not uses 3D information.

We also found manually small subsubsets of domains that are wrongly clustered
by  USM, but that SSEARCH classifies almost perfectly, using the
same clustering algorithm used in \citet{Pelta1}. Due to paper limitations, they are not shown here but in the web page {\tt dmi.uib.es/people/jairo/USM/}.
In the fist one in the web, for instance, there are proteins that belong to
two different classes (the wider tree level) that are wrongly clustered 
by USM.  
\section{Conclusion}

The USM based method under a simple contact map format clusters poorly
protein groups easily discriminable and, in general,
their clusters are father form CATH than the ones generated by some methods
based on sequences;    the three small real sets that it  clusters almost correctly
are by no means an evidence that the method is robust for clustering. 
In addition, the method is very far from
becoming a reliable protein comparison method from the point of view
of deciding if two given protein structures are similar or not,
one of the most important procedures to detect protein function similarity.

\section*{Acknowledgements}
We thank the following people: Kevin Karplus for bringing out the idea on the assignment of lengths to the 
tree edges in order to find how far a distance is from an additive one;
Ming Li for  discussing with the authors the first version of this paper, and
 Joe Mir\'o for his comments on an early draft.
We acknowledge the financial support of the Spanish and
UE grant BFM2003-00771 and the UE grant INTAS IT 04-77-7178.


\end{document}